\documentclass[twocolumn,showpacs,preprintnumbers,amsmath,amssymb]{revtex4}
\usepackage{graphicx,psfrag,bbm,latexsym,color,dcolumn,bm,dsfont,bbm,color,mathrsfs,bbold,latexsym,amsmath,amsfonts,amssymb,epsfig}

\newcommand{\beq}{\begin{equation}}
\newcommand{\eeq}{\end{equation}}

\newcommand{\beqa}{\begin{eqnarray}}
\newcommand{\eeqa}{\end{eqnarray}}

\begin{document}

\title{Lorentz Covariance and the Dimensional Crossover of 2d-Antiferromagnets}
\author{P.R. Crompton}
\affiliation{Institut f\"ur Theoretische Physik I., Universit\"at Hamburg, D-20355 Hamburg, Germany.}
\affiliation{Center for Theoretical Physics, Massachusetts Institute of Technology, Cambridge, MA  02139, USA.}
\vspace{0.2in}
\date{\today}

\begin{abstract}
{We derive a lattice $\beta$-function for the 2d-Antiferromagnetic Heisenberg model, which allows the lattice interaction 
couplings of the nonperturbative Quantum Monte Carlo vacuum to be related directly to the zero-temperature fixed points of 
the nonlinear sigma model in the presence of strong interplanar and spin anisotropies. In addition to the usual 
renormalization of the gapful disordered state in the vicinity of the quantum critical point, we show that this leads to a 
chiral doubling of the spectra of excited states.} 
\end{abstract}


\maketitle

The 2d-Antiferromagnetic Heisenberg model is a well studied system that has two RG fixed points that have been identified 
at zero temperature through various applications of the 2d nonlinear sigma model \cite{sig,hsig}. Dependening on 
the relative anisotropy of the exchange coupling, $J$, between the two spatial directions of the system, the groundstate 
of the 2d antiferromagnetic groundstate is found to be either N\'eel ordered and gapless, or a gapful quantum disordered 
state. Varying the anisotropy drives dynamical fluctuations causing a zero temperature phase transition between the two 
groundstates \cite{sand}, whereby the system effectively crosses over from 2d to 3d \cite{dimc}. Without the inclusion of
 a $\theta$-term, the 2d nonlinear sigma model is used to give an effectively classical description of the N\'eel ordered 
groundstate, and the quantum nature of the disordered state arises in some nontrivial way through the dynamical scale 
evolution of the system. It has been questioned whether the inclusion of an explicit source term for quantum fluctuations 
would be of relevance in the N\'eel phase \cite{2d,qc}, in order to understand the mechanism of symmetry breaking, but 
the inclusion of such a $\theta$-term provides an irrelevant perturbation. However, if no source term for quantum 
fluctuations is included, the renormalization scale of the system can only be defined through phenomenological input. 
This picture has been successfully verified in detail by comparing nonlinear sigma model predictions for the scaling of 
the correlation length in the N\'eel, and quantum disordered regimes, with numerics obtained from Quantum Monte Carlo (QMC) studies 
\cite{tro,ani}. 

Recently a treatment of the reverse picture has been given via a conformal analysis of quantum spin chains \cite{muss}.
Importantly, this identifies the effect of dynamical fluctuations on the $\theta$-term. In principle, this effect should 
simply be to rescale the couplings of the underlying sine-Gordon model by a dynamical factor. However, we have recently 
pointed out that the convergence of the perturbative deformation is not guaranteed and a nonperturbative 
renormalization prescription is required \cite{me-jstat}. In this work, we extend this nonperturbative renormalization 
picture to give a description of the role of quantum effects in the dynamical fluctuations of the 2d-Antiferromagnetic 
Heisenberg model. 
Whether or not a $\theta$-term is relevant in the N\'eel phase \cite{troe}, the O(3) spin operators of 
the nonperturbative QMC method \cite{qmc} have already been obtained through a form of topological dimensional reduction. 
Consequently, there are IR cutoff effects associated with the finite lattice system dynamics, and it is difficult to 
disentangle these effects from genuine mechanisms of symmetry breaking. Recent treatments based around the 2d O(3) model 
have included the effect of a marginally irrelevant topological term possessing a U(1) symmetry \cite{qc}. Our new 
nonperturbative approach extends this picture by considering the effect of a finite IR lattice cutoff within the couplings 
of these terms.

We focus on the nonperturbative properties of the continuous-time QMC method. This scheme has the same basic transfer 
matrix structure that is defined for the Density Matrix Renormalization Group (DMRG) method \cite{DMRG1,DMRG2}, 
but the numerical basis of the lattice partition function for the continuous-time QMC method is defined through the application of 
classical loop-cluster methods. There are a number of closely related QMC schemes that can also be used to 
generate this lattice partition function, and these are discussed in detail in the review in \cite{Evertz}. However, we 
are single out the continuous-time scheme because the loop-cluster Monte Carlo updating process is slightly 
different. The closed loops in this scheme, which represent the trace of the lattice partition function, are generated 
through a sequence of successive local Monte Carlo updating decisions, and this is in contrast with other QMC schemes, 
closer to the original classical loop-cluster method, where the Monte Carlo updating decision is made by comparing the 
probabilistic weights of closed loops. All of the QMC schemes which use the DMRG form of the transfer matrix in 
\cite{DMRG1,DMRG2} can 
be given the same interpretation as systems of O(3) spin vectors, but the continuous-time method has the addtional feature 
that the lattice partition function is defined to be analytically continuous in Euclidean-time over some finite interval. 
The reason why this is important is because, unlike the dynamical scaling relations of classical spin systems \cite{z}, 
the dynamical critical exponent for the QMC Euclidean-time direction only relates to a topologically complete system in 
the large lattice volume and stochastic probability distribution limits of the QMC method \cite{PRD}. Therefore, 
a complete understanding of numerical correction effects and a direct comparison with the couplings of the 2d nonlinear 
sigma model is difficult with the generic method because finite-size lattice systems only asymptotically 
approach these limits. Using the special continuous analytic property of the continous-time methods we are now able to
define this limiting process in a new way. The nonperturbative lattice renormalization group equation for the 
2d-Antiferromagnetic Heisenberg model that, we now propose, is useful as it can save computational effort in numerical 
studies by interpolation of the dynamical scaling. 

\section{Quantum Monte Carlo method}

In order to understand the renormalization group flow of the lattice couplings in the QMC method that are 
defined using the transfer matrices in \cite{DMRG1,DMRG2}, and how these relate to the 2d nonlinear sigma model, it is important to 
understand why these methods do not, in general, yield Lorentz invariant systems. This particular form of the QMC method 
has a deceptively simple overlap with the 2d O(3) model. Numerical loop evolution generates the lattice path integrals and 
partition functions in the generic approach. This loop evolution proceeds over both the spatial and Euclidean-time extents 
of the lattice on an equal footing through the evaluation of local Monte Carlo decisions. The three component O(3) quantum 
spin operators are represented by their $S_{z}$-component as a discrete spin defined on the spatial lattice sites, and the 
$S_{x}$ and $S_y$ components of the spin operators are defined through the projection of the probability distribution of 
the loop evolution defined for the Euclidean-time extent \cite{DThe}. Thus we arrive at a very similar picture to the 
continuum 2d O(3) model with a $\theta$-term, but the crucial difference is that the topological term has a finite IR 
cutoff. Although for sufficiently large lattice volumes the distributions of the projections of the operators onto the 
spatial and Euclidean-time lattice directions do converge, these distributions are not constrained to be identical in the 
lattice ensemble through the definitions of the loop evolution \cite{clus}. Although the transfer matrix is isotropic in 
space and Euclidean-time, the individual loops that are realised through probabilistic decisions are not. Spatial and 
Euclidean-time isotropy is only realised in the stochastic limit of the lattice ensemble. In practice, a model defined 
with isotropic interaction couplings can be realised as being locally anisotropic through the dynamics.

Our aim in this article is to attempt to quantify the effect of this local anisotropy. In general, if we focus just on the 
spatial properties of the lattice ensemble then the critical behaviour, measured via the probability distribution of the 
$S_{z}$-component of spin, can be defined as a function of the lattice interaction coupling, $J$. This follows from the 
usual finite-size scaling (FSS) picture, which consists of asymptotic expansion about renormalization group fixed points 
\cite{FSS}. However, since the generic QMC spin operators are also defined through a form of topological dimensional 
reduction, this implies that the fixed point can also be approached smoothly in $\beta$ (the inverse temperature and 
cutoff scale for Euclidean-time). If this equivalence property is realised, the dynamical critical exponent is necessarily 
unity and the system is Lorentz invariant. However, in general, this scaling picture is disrupted by IR cutoff effects, 
and consequently the numerically realised system cannot be Wick-rotated. Therefore by defining an exact basis of dynamical 
fluctuations, the equivalence with the O(3) spin vectors is broken in the generic QMC method that is defined through the 
transfer matrix in \cite{DMRG1,DMRG2}.

What we will now therefore do is simply change the emphasis of the analysis. We will start with a Wick-rotated definition 
of the loop operators of the QMC method, to make the O(3) equivalence exact, and then consider the IR cutoff effects that 
arise in finite-size lattice system in this choice of basis. This will enable us to quantify the breakdown of Lorentz 
invariance through the emergence of dynamical scale effects, and to quantify the effect of dynamical fluctuations on the 
irrelevant quantum perturbations of the 2d-Antiferromagnetic Heisenberg model through nonperturbative renormalization. 
We do this for the continuous-time QMC method in \cite{qmc}, because the lattice partition function is analytically continuous 
in Euclidean-time upto some finite scale.

\section{Wick rotation}
The new approach we take is to identify an analytic analogue of the QMC partition function, which is Wick-rotated. We 
express the O(3) spin operators using a formalism developed to treat the probabilistic dynamics of lattice systems 
analytically \cite{pres}. This gives an exact description of the lattice ensemble vacuum in the vicinity of the large 
volume stochastic limit, i.e. in the UV. We then expand this description towards the IR. The asymptotic freedom of the 2d 
O(3) model has been known for some time \cite{poly}. It is also known that the IR fixed point associated with the Berry 
phase can be treated via perturbative deformations \cite{fat}. What we treat here is the most general form of deformation 
of the irrelevant Berry term, via couplings which are modified by the dynamical fluctuations, where the convergence of 
the perturbative expansion would not then be guaranteed.

In general, a nonperturbative lattice ensemble is only known through the expectation of its numerical matrix elements 
$\lambda$. These matrix elements can be defined in terms of the generalised spin vectors $\bm{n}$ that describe the state 
of the spins on the lattice. For the QMC method, these matrix elements are defined in terms of both $S_z$ components, 
defined on the spatial lattice sites, and also $S_x$ and $S_y$ components, defined through the projection of the 
probability distribution onto the Euclidean-time extent of the lattice. In the stochastic limit we expect the system to be 
Lorentz invariant, therefore we can define the Wick-rotation of the lattice system through the analytic continuation 
$J\equiv i\theta$. The spin vectors defined on the spatial lattice sites will then represent the orientation of the $S_x$ 
and $S_y$ components of spin, and the spin vectors defined on the Euclidean-time direction will represent the projection 
of the probability distribution defined for the $S_z$ component of the spins. The reason for doing this is that in the 
continuous-time QMC method \cite{qmc} the number of lattice sites in the Euclidean-time direction is variable. Therefore, 
if we Wick-rotate the definitions we are able to consider the FSS of the topological terms and the IR cutoff effects 
in quantifiable lattice units. 

The transfer matrix of the 2d AFM is a $8\times8$ matrix \cite{DMRG}, and so define the lattice matrix elements through 
three projection indices: onto the Euclidean-time direction, and onto the two separate spatial directions. The implicit 
isotropy between the $S_x$ and $S_y$ components of spin defines a locally conserved current, $\theta$, associated with 
each spatial lattice site. We therefore define two separate matrix elements for the two different spatial directions on 
the lattice, 

\beqa
\label{Hilberta}
A(\bm{n}) \! \equiv \!\!\!\!\!\!\!\!
\sum_{(s,s',s'')}^{T \otimes \Theta_{1} \otimes \Theta_{2}} \!\! \sum_{\sigma \in G}
\lambda_{ss's''\sigma}(\bm{n})\frac
{\langle \bm{n}\oplus \bm{1}_{s\sigma} \oplus \bm{1}_{s'\sigma}\oplus \bm{1}_{s''\sigma}| \theta_{1} \rangle}
{\langle \bm{n}| \theta_{1} \rangle}\, \nonumber \\
\eeqa
\beqa
\label{Hilbertb}
B(\bm{n}) \! \equiv \!\!\!\!\!\!\!\! 
\sum_{(s,s',s'')}^{T \otimes \Theta_{1} \otimes \Theta_{2}} \!\! \sum_{\sigma \in G}
\lambda_{ss's''\sigma}(\bm{n})\frac
{\langle \bm{n}\oplus \bm{1}_{s\sigma} \oplus \bm{1}_{s'\sigma}\oplus \bm{1}_{s'' \sigma}| \theta_{2} \rangle}
{\langle \bm{n}| \theta_{2} \rangle}\, \nonumber \\
\eeqa

where $G$ is the discrete $Z(2)$ algebra of the $S_z$ spins, $\sigma$ is an element of $G$, 
$\,\Theta_{1},\Theta_{2}\equiv L\,$, and $T$ is the Trotter number of the Euclidean-time extent of the lattice.
The partition function for the 2d-Antiferromagnetic Heisenberg model is then given, in terms of these matrix elements, 
as a path-integral over these two conserved currents, 

\beqa
\label{partitionz}
\mathcal{Z} = \int {\rm{d}} [\theta_{1}]  {\rm{d}} [\theta_{2}] 
\,\,\, {\rm{exp}} \Big[ \,\int_{0}^{\beta} \!\!\!\!\!\!\!\!\!\!\!\!\! & & A(\bm{n}_{s})\,i\theta_{1} + B(\bm{n}_{s})\,i\theta_{2}\nonumber \\  & - & V(\bm{n}_{s})
\,\,ds \,\Big]
\eeqa

where $V$ comprises the off-diagonal contributions to the spin operator basis, which is simply a general matrix 
element on the $T\otimes\Theta_{1}\otimes\Theta_{2}$ lattice with no special symmetries,

\beqa
\label{Hilbertv}
V(\bm{n}) \!\! & \equiv & \!\!\!\!\!\!\!\!
\sum_{(s,s',s'')}^{T \otimes \Theta_{1} \otimes \Theta_{2}} \!\! \sum_{\sigma \in G}
\lambda_{ss's''\sigma}(\bm{n})\frac
{\langle \bm{n}\oplus \bm{1}_{s\sigma} \oplus \bm{1}_{s' \sigma}\oplus \bm{1}_{s'' \sigma}| \bm{n}
\rangle}
{\langle \bm{n}| \bm{n} \rangle}\, \nonumber \\
\eeqa

Instead of a single compact Berry phase term, in this nonperturbatively-motivated operator formalism we have two separate 
source terms in $\theta$, one for each of the two anisotropic spatial directions. There is also an implicit cross-term in 
$\theta_{1}\otimes\theta_{2}$, which comes from $V$. From this latter term we are able to generate local states like the 
plaquette ordered groundstates considered in \cite{plaq} because of the four-fold symmetry coming from the relative signs 
of $\theta_{1}$ and $\theta_{2}$. However, we can also consider what happens if we are unable to continue either 
$\theta_{1}$ or $\theta_{2}$ through $2\pi$ because of the finite IR cutoff. There will still be an equal number of 
instantons and anti-instantons within the vacuum, conserving the total topological charge, but if the rotational symmetry 
about the $S_x$ and $S_y$ spin component plane is lost, then the four-fold parity symmetry is broken. This symmetry is 
broken down to a two-fold symmetry such that the instantons are no longer symmetric under reflection about the $S_{z}$ 
component of spin and their chirality is lost. In the 2d O(3) model the N\'eel phase is defined by having a larger 
correlation length than that of the dynamical fluctuations. The same is true of the groundstate realised by the generic 
QMC method, but at finite Trotter number, the same is not necessarily true of correlations in the $4\pi$ rotational 
symmetry of the instantons away from the limit in which the lattice ensemble is Lorentz invariant. 

To simplify the cross-term of $\theta_{1}\otimes\theta_{2}$, we can define the projection of the $\theta_{1}$ component 
of $V$,

\beqa
\label{Hilbertc}
C(\bm{n}) & \!\! \equiv \!\! & \langle \bm{n} | V^{2} ( \bm{n}_{s'} ) | \bm{n} \rangle = \!\!\!\!\!\!
\sum_{(s,s',s'')}^{T \otimes \Theta_{1} \otimes \Theta_{2} } \!\! \sum_{\sigma \in G}
\lambda_{s s's''\sigma}(\bm{n}) \langle \bm{1}_{\sigma s'}| \bm{n} \rangle \nonumber \\ & & 
\!\!\!\!\!\!\! \!\!\!\!\!\!\!\! \!\!\!\!\!\!\!\! \times\!\!\!
\sum_{(s,s',s'')}^{T \otimes \Theta_{1} \otimes \Theta_{2} } \!\! \sum_{\sigma \in G}
\lambda_{s s's''\sigma}(\bm{n})
\langle \bm{n}\oplus \bm{1}_{s\sigma} \oplus \bm{1}_{s'\sigma} \oplus \bm{1}_{\sigma s''}| \bm{n}_{s'} 
\rangle 
\eeqa
Substituting, this then yields the partition function,

\beqa
\mathcal{Z} = \int {\rm{d}} [\theta_{1}] {\rm{d}} [\theta_{2}] \,\,\, {\rm{exp}} \Big[ & & \!\!\!\!\!
\int_{0}^{\beta} A(\bm{n}_s)i\theta_{1} + \sqrt{C(\bm{n}_s)} i\theta_{1} \nonumber \\ & &  \!\!\!\!\!\!\!\!\!\!\!\! + \, B(\bm{n}_s)\,i\theta_{2} \, - \, V'(\bm{n}_s) \,\,ds \,\Big] 
\eeqa
where, $V'(\bm{n}_s) \equiv V(\bm{n}_s) -  C(\bm{n}_s)$.

\section{lattice spacing}

Our motivation for making a Wick-rotation is to quantify the effect of dynamical fluctuations on the 
irrelevant Berry term in the 2d-Antiferromagnetic Heisenberg model. In the above new Wick-rotated operator definitions, 
$\theta_{1}$ and $\theta_{2}$ are noncompact abelian variables, and the Euclidean-time extent is used to represent the 
multiplicity of these phases. Consequently, each of the discrete intervals of the Euclidean-time extent now corresponds to 
a different $\theta$-vacuum. To relate this new formalism to the usual compact definition of topological charge (given for 
the 2d O(3) model) we must somehow select one of these $\theta$-vacua over the others to be our reference compact sector. 
Conveniently, this choice of $\theta$-vacuum naturally arises from the nonperturbative dynamics in the form of the 
numerical expectation value that is realised by the projection of the probability distribution defined for 
Euclidean-time \cite{me-jstat}. Stating this simply, each spin site on the $T\otimes\Theta_{1}\otimes\Theta_{2}$ lattice 
has an associated matrix element $\lambda$, and the Euclidean-time projection of $\lambda$ has a larger value in one of 
the discrete intervals of the Euclidean-time extent than in the others. This is, however, a spatial site-specific result. 
In principle, each spatial site can have the maxima of this projection arise in a different Euclidean-time interval. 
Different local sites can, therefore, correspond to different $\theta$-vacua, which presents a source of nonintegrable 
singularity in analytically continuing the lattice interaction coupling, $J$ \cite{new-prd}. These singularities, though, 
are precisely the form of dynamical fluctuation-induced IR cutoff effect that we are aiming to quantify.

To quantify these properties, we introduce local measures of lattice spacing, $a$ and $b$, defined in units of $\theta$ 
and $\beta$, respectively. These are given as the difference between the projection of the matrix element realised on a 
given lattice site and the expectation of the projection averaged over the corresponding lattice extent,

\beqa
\theta_{1} \, a_{s'} & \equiv & A(\bm{n}_{s'}) -\langle A(\bm{n}_{s'}) \rangle_{\Theta_{1}} \nonumber \\
\beta  \, a_{s} & \equiv & A(\bm{n}_{s}) - \langle A(\bm{n}_{s}) \rangle_{T} \nonumber \\ 
\,\theta_{2} \, b_{s''} & \equiv & B(\bm{n}_{s''}) -\langle B(\bm{n}_{s''}) \rangle_{\Theta_{2}}\nonumber \\ 
\beta  \, b_{s} & \equiv & B(\bm{n}_{s}) - \langle B(\bm{n}_{s}) \rangle_{T} \nonumber \\
\eeqa
Practically, $A(\bm{n}_{s})$ and $B(\bm{n}_{s})$ can be found directly as they correspond to the diagonal entries of the 
numerical transfer matrix of the continuous-time QMC method. Similarly, the two operator projections $A(\bm{n}_{s'})$ and 
$B(\bm{n}_{s''})$ can be calculated by summing the local continuous-time QMC transfer matrix entries over the spatial 
sites rather than Euclidean-time. Thus, in a formal mathematical sense these lattice spacing definitions define the 
support of the matrix elements found by integrating the local $T\otimes\Theta_{1}\otimes\Theta_{2}$ matrix elements 
over a given lattice direction.

A related quantity is the parallel transport between neighbouring lattice sites for which we can define the 
following local derivative operators, 
\beqa 
\partial_{s} A(\bm{n}_{s}) & \equiv & A(\bm{n}_{s+1}) - A(\bm{n}_{s}) = i\theta_{1} A(\bm{n}_{s}) \nonumber \\ 
\partial_{s} B(\bm{n}_{s}) & \equiv & B(\bm{n}_{s+1}) - B(\bm{n}_{s}) = i\theta_{2} B(\bm{n}_{s}) 
\eeqa

\section{lattice $\beta$-function}

To identify a new nonperturbative lattice $\beta$-function we define an effective action for the partition function of 
(6), which is given as an expansion in the self-energy terms of the dynamical basis.  A general loop generated by the 
Monte Carlo process is of the form $T\otimes\Theta_{1}\otimes\Theta_{2}$ and is given by the action of 
$V(\bm{n})$-operators on the vacuum. Similarly, the action of the $A(\bm{n})$-operators generates loops of the form 
$T\otimes\Theta_{1}$, and the $B(\bm{n})$-operators generate loops of the form $T\otimes\Theta_{2}$. Thus, the 
self-energy terms of the Wick-rotated dynamical basis we have defined, which describe the projection of the $A$ and $B$ 
operators onto $V$, are of the form $\Theta_{1}$ and $\Theta_{2}$, and correspond to the $\theta$-symmetry breaking 
component of the vacuum. The effective action is of the form, 

\beq
S=S_1 + S_2 + S_{12}
\eeq
where,
\beqa
S_1  \! & = & \! \int_{0}^{\beta} ds \,\,\partial_{s} A(\bm{n}_{s}) - \partial_{s} [V'(\bm{n}_{s}) A(\bm{n}_{s})] -V'(\bm{n}_{s}) \nonumber \\
S_2  \! & = & \! \int_{0}^{\beta} ds \,\,\partial_{s} B(\bm{n}_{s}) - \partial_{s} [V'(\bm{n}_{s}) B(\bm{n}_{s})] -V'(\bm{n}_{s}) \nonumber \\
S_{12}  \! & = & \! \int_{0}^{\beta} ds \,\, \partial_{s} \! \sqrt{C(\bm{n}_{s})} - \partial_{s} [V'(\bm{n}_{s}) \sqrt{C(\bm{n}_{s})}] +V'(\bm{n}_{s})\nonumber \\ && \quad\quad\quad\quad  
\eeqa

The terms of the effective action are of the same form as the undeformed action in (\ref{partitionz}). These terms are 
defined through a form of topological dimensional reduction where an explicit integration is performed over $\beta$ to 
project out the Euclidean-time dependence. Next we integrate out the effective action terms in a second topological 
dimensional reduction step by integrating out the expansion over the compact sectors that define each lattice matrix 
element. The purpose of this second step is to make the effective action compact, which allows us to then make a direct 
comparison between our new RG flow and that of the 2d nonlinear sigma model. To do this we introduce two new variables; 
$x$ which is a general position index on $T\otimes\Theta_{1}$, and $y$ which is a general position index on 
$T\otimes\Theta_{2}$. 

Our aim, following Ref.\onlinecite{muss}, is to treat the effect of irrelevant quantum fluctuations on the relevant dynamical 
fluctuations of the 2d nonlinear sigma model, given in Ref.\onlinecite{sig}. Our new operator formalism is defined to be exact 
in the basis of these dynamical fluctuations from the properties of the continuous-time QMC method. In practice, the 
numerics suffer from IR cutoff effects but we have Wick-rotated the operator definitions in order to quantify this effect 
on the topology of the O(3) spin vectors. In principle, the IR cutoff implies that the $2\pi$-rotational symmetry of the 
$S_x$ and $S_y$ plane components is broken. However, we have the freedom, from the renormalization group properties of the 
relevant dynamical fluctuations, simply to rescale the couplings of the operators such that the $2\pi$-rotational 
symmetry is present. There is then a new singularity that sits on the lattice vertices, which appears essentially by 
modifying a finite number of the poles of the lattice system to branch points.

The second topological dimensional reduction step is straightforward to evaluate for the first two terms in (9) 
using the local lattice spacing definitions given in (7).
\beqa
\label{eff1}
S_{1} & = & \!\!\! \int_{0}^{\beta} \!\! \int_{-\pi}^{\pi}  d^{2}x \quad a_{s'}\,\partial^{2}_{x} A(\bm{n}) - a_{s}\,\partial^{2}_{x} V'(\bm{n})   - 
a_{s}\wedge a_{s'}\nonumber \\ \\
\label{eff2}
S_{2} & = & \!\!\! \int_{0}^{\beta} \!\! \int_{-\pi}^{\pi}  d^{2}y \quad b_{s''}\,\partial^{2}_{y} B(\bm{n}) - b_{s}\,\partial^{2}_{y} V'(\bm{n})   - 
b_{s}\wedge b_{s''}\nonumber \\
\eeqa
These isotropic terms are of the form of the corresponding sine-Gordon model description of quantum spin chains. This is 
expected since the interaction between the two spatial lattice directions is ignored. These terms imply, following 
Ref.\onlinecite{muss}, that the effect of quantum fluctuations can be simply rescaled the into the existing couplings, which 
describe the dynamical fluctuations. The cross-term, which links the two spatial lattice directions, is slightly more 
involved, however, since it involves a choice of branch.

\beqa
\label{eff3}
S_{12} \! & = & \! \int_{0}^{\beta} \! \int_{-\pi}^{\pi} d^{2}x \,\, - [ 1 - \langle \!\sqrt{C(\bm{n_{s}})}\rangle_{\Theta_{1}}]\, \partial^{2}_{x} V'(\bm{n}) 
\nonumber \\ && +\,(1 + b_{s}\wedge b_{s''}) [ 1 - \langle V'(\bm{n_{s}})\rangle_{\Theta_{1}} ]\,\partial^{2}_{x}[  \sqrt{C(\bm{n})  } ] \nonumber \\
&& -\,\partial_{x} \langle \!\sqrt{C(\bm{n}_{s})}\rangle_{\Theta_{1}}\partial_{x} 
\langle V'(\bm{n}_{s})\rangle_{\Theta_{1}} \nonumber \\ \nonumber \\ 
& = &  \int_{0}^{\beta} \! \int_{-\pi}^{\pi}  d^{2}x \quad a_{s'}\wedge (1 + b_{s}\wedge b_{s''})\,\partial^{2}_{x} [\sqrt{C(\bm{n})}]\nonumber \\  && - a_{s}\wedge (b_{s}\wedge b_{s''}) \,\partial^{2}_{x} V'(\bm{n})   -  (a_{s}\wedge a_{s'})\wedge (b_{s}\wedge b_{s''}) \nonumber \\ 
\eeqa
The cross-term contribution to the effective action, within the contour between $\pi$ and $-\pi$, is of the same form as
 (\ref{eff1}), but the contribution from the branch point on the boundary leads to an additional phase contribution to 
the action \cite{PRD}. By comparing (\ref{eff1}) and (\ref{eff3}) we find that the cross-term contribution is 
implicitly anisotropic and, therefore is of the form of the doubel Sine-Gordon model not the sine-Gordon model \cite{muss}. 
This means that although the effect of quantum fluctuations on the dynamical fluctuations is irrelevant (amounting simply 
to a change in the renormalization scale of the vacuum), the effect of quantum fluctuations is relevant for the Lorentz 
covariance scale of the dynamical fluctuations. The effect of having a branch point associated with parity in the 
cross-term, is that the parity of the vacuum (that allows an exact symmetry between the instantons and anti-instantons) 
can be explicitly broken via quantum fluctuations. It is only in the limit when $b_{s}\wedge b_{s''}=1$ that the effective 
action term in (\ref{eff3}) is Lorentz invariant and this effect is vanishing.  

gap above the ordered groundstate. What we should therefore find is that the spectrum of excited states exhibits a doubling due to the 
degeneracy of the broken chirality of the instanton anti-instanton pairs, when we are in the region of the classical dimensional crossover 
of the 2d quantum antiferromagnet with the presence of strong anisotropies due to these quantum fluctuation effects.

The sine-Gordon renormalization group equations for (\ref{eff1}) and (\ref{eff2}) are then trivially modified in 
the presence of anisotropy \cite{kost}. However, this leads to the emergence of a new unstable fixed point and finite 
renormalized region in the vicinity of the second order quantum critical point.

\beqa
\label{RG}
\frac{da_{s}}{dl} & = & -\frac{1}{2} a_{s}^{2}\left(\frac{a_{s'}-b_{s}\wedge b_{s''}}{\pi}\right)^{2} \\ \nonumber
\frac{da_{s'}}{dl}& =& (a_{s'}-b_{s}\wedge b_{s''})(2-2a_{s})
\eeqa

The axes-crossing in the usual hyperboloid scaling picture of the $(a_{s},a_{s'})$ phase plane is simply shifted by 
the inclusion of branch singularities. This leads to a very simple analytic rescaling of the relevant couplings of the 
gap state, but it is one which is not easily quantified perturbatively in terms of a global cross-term (as in \cite{sig}) 
because of the local site dependence of the branch \cite{PRD}. It was argued in \cite{bas} that such a contribution should 
be responsible for the appearance of deconfined spinons at a finite energy gap above the ordered groundstate. What we 
should, therefore, find is that the spectrum of excited states exhibits a doubling due to the degeneracy of the broken 
chirality of the instanton/anti-instanton pairs when we are in the region of the classical dimensional crossover of the 2d 
quantum antiferromagnet through the Lorentz covariance of the irrelevant quantum fluctuations.
\section{summary}

From early spinwave analyses of the 2d quantum antiferromagnet it was concluded that the role of quantum fluctuations is 
irrelevant to the stability of the N\'eel ordered groundstate \cite{and}. Subsequent refinement of these arguments has 
suggested that the crucial analytic property of the analysis, which enables this stability, is the continuity of 
generalised spin operators that describe the vacuum \cite{poly2,2d}. In this article, we have now considered the role 
of dynamical cutoff effects on these quantum fluctuations in an exact basis of dynamical fluctuations following the 
nonlinear sigma model treatment given in Ref.\onlinecite{sig}. The crucial difference from previous studies is that we have now 
considered the role of anisotropy on modifying the renormalization cutoff scale on the phenomenological couplings of the 
ordered state. Whilst it is known to high accuracy that the N\'eel ordered groundstate closely follows spinwave theory 
predictions \cite{sand2}, considerably less is known about relevant renormalization of the dynamical couplings
by quantum fluctuations when the Lorentz symmetry of the system is broken. This nonperturbative scaling treatment forms a 
more realistic picture of the numerical scaling of the dynamical basis defined by the generic QMC method, which is only 
truly Lorentz invariant asymptotically close to the stochastic limit. Similarly, the new nonperturbative renormalization 
group formalism provides a loop expansion formalism which is suitable for rigorously probing the conformal correspondence 
of experimental high temperature superconductivity data at strong interaction couplings. Our main new result from this 
analysis is that the effect of the branch points that describe the chirality of the instantons can be spontaneously broken 
through quantum fluctuations through the Lorentz covariance of the couplings. We have argued, following \cite{bas}, that 
this should then lead to a doubling of the spectrum of excited states in the vicinity of the quantum critical point of the 
2d quantum antiferromagnet when the dynamical couplings are strong.

\end{document}